\documentclass[prb,twocolumn,amsmath,amssymb,showpacs,superscriptaddress]{revtex4}

\usepackage{graphicx}

\begin{document}

\newcommand{\etal}{{\it et al.}\/} 
\newcommand{\gtwid}{\mathrel{\raise.3ex\hbox{$>$\kern-.75em\lower1ex\hbox{$\sim$}}}} 
\newcommand{\ltwid}{\mathrel{\raise.3ex\hbox{$<$\kern-.75em\lower1ex\hbox{$\sim$}}}}

\title{Neutron Scattering Resonance and the Fe-pnictide Superconducting Gap}

\author{T.A.~Maier} \email{maierta@ornl.gov} \affiliation{Center for Nanophase Materials Sciences, Oak Ridge National Laboratory, Oak Ridge, TN 37831-6164} \affiliation{Computer Science and Mathematics Division, Oak Ridge National Laboratory, Oak Ridge, TN 37831-6494}

\author{S. Graser} \email{graser@phys.ufl.edu} \affiliation{Center for Electronic Correlations and Magnetism, Institute of Physics, University of Augsburg, D-86135 Augsburg, Germany}

\author{D.J. Scalapino}\email{djs@physics.ucsb.edu} \affiliation{Department of Physics, University of California, Santa Barbara, CA 93106-9530 USA}

\author{P. Hirschfeld} \email{pjh@phys.ufl.edu} \affiliation{Department of Physics, University of Florida, Gainesville, FL 32611, U.S.A.}

\date{\today} 
\begin{abstract}
	
	The existence of a neutron scattering resonance at a wavevector $q^*$ implies a sign change of the gap between two Fermi surface regions separated by wavevector $q^*$. For the Fe pnictides, a resonance has been observed for a wavevector $q^*$ which connects a hole Fermi surface around the $\Gamma$ point with an electron Fermi surface around the $X$ or $Y$ points of the 1 Fe/unit cell Brillouin zone. Here we study the neutron scattering resonance for a five orbital model within an RPA-BCS approximation. Our results show that both sign-switched and extended s-wave gaps are consistent with the present data for $q^*$ near $(\pi,0)$ and that scattering at other momentum transfers can be useful in distinguishing between gap structures.
\end{abstract}

%\pacs{71.10.Fd, 03.75.Ss, 74.25.Ha }
\maketitle

%\section{Introduction}
Recent neutron scattering studies~\cite{ref:christianson,ref:lumsden,ref:li} find a resonance in the superconducting state of the 122 Fe pnictides which appears below $T_c$ and has a wave vector $q^*\sim(\pi,0)$ in the unfolded (1 Fe/cell) Brillouin zone. One would like to understand what this resonance implies about the structure of the superconducting gap $\Delta(k)$. As is well known, the occurrence of resonances in the neutron scattering spectrum depends through the BCS coherence factors on the relative signs of the gap on different parts on the Fermi surface separated by $q$, and thus gives insight into the momentum structure of the superconducting gap. Nuclear resonance Knight shift measurements~\cite{ref:1,ref:2,ref:3} support a singlet pairing state and $A_{1g}$ and $B_{1g}$ gaps have been found in various fluctuation-exchange calculations~\cite{ref:4,ref:5,ref:6,ref:7} as well as renormalization group studies.~\cite{ref:8,ref:9} Motivated by the experiments and results from these calculations, we propose to examine the relationship of the resonance to its $q^*$ space location and the $k$-dependence of the gap. Previous calculations~\cite{ref:10,ref:11} have found that a resonance occurs at a wave vector $q^*\sim(\pi,0)$ for a sign switched~\cite{ref:12} $s_\pm$ gap. This gap has a sign change between the inner hole Fermi surfaces that surround the $\Gamma$ point and the electron Fermi surfaces around the $X(\pi,0)$ and $Y(0,\pi)$ points of the unfolded (1 Fe/cell) Brillouin zone. One of these calculations was for a two-orbital model\cite{ref:10} and the other used a four-band model but neglected the role of orbital-band matrix elements.~\cite{ref:11} To adequately describe the region of the Fermi surface of the Fe-pnictides one needs at least three orbitals~\cite{ref:13} and the orbital-band matrix elements are known to play an essential role in determining the $q$ dependence of the magnetic susceptibility.~\cite{ref:7} Here we will use an RPA-BCS approximation~\cite{ref:14} to calculate the neutron scattering response for a five-orbital tight binding model with onsite Coulomb and exchange interactions. This model has sufficiently many orbitals to describe the electronic structure in the relevant regions near the Fermi energy. We will also take account of the orbital-band matrix elements.

\begin{figure}
	[htbp]
	\includegraphics[width=3in,clip,angle=0]{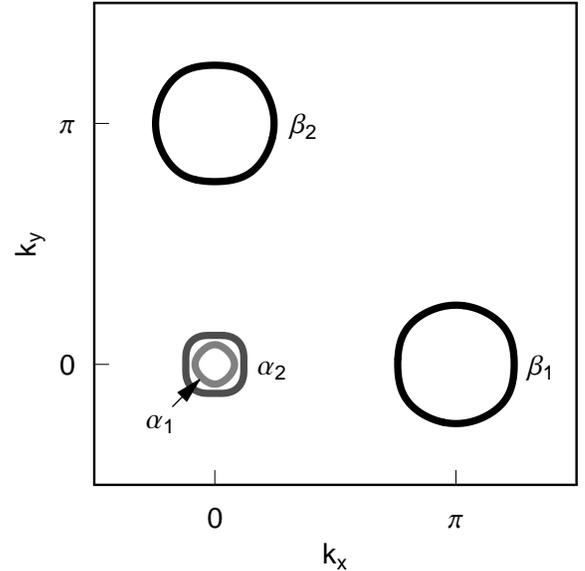}
	\caption{Fermi surfaces for the 5-orbital model at a doping $x=0.125$. There are two hole Fermi surfaces $\alpha_1$ and $\alpha_2$ around the $\Gamma$ point and two electron Fermi surfaces $\beta_1$ and $\beta_2$ around the $X(\pi,0)$ and $Y(0,\pi)$ points, respectively.} \label{fig:1} 
\end{figure}

Figure~\ref{fig:1} shows the Fermi surfaces for this 5-orbital model with tight binding parameters fit to reproduce the LDA bandstructure~\cite{ref:15} near the Fermi energy. 
For a doping $x=0.125$, there are two hole Fermi surfaces $\alpha_1$ and $\alpha_2$ around the $\Gamma$ point of the unfolded (1 Fe/cell) Brillouin zone and two electron Fermi surfaces $\beta_1$ and $\beta_2$ around the $X$ and $Y$ points. For certain ranges of interaction parameters, the leading pairing instability in an RPA fluctuation exchange approximation occurs for an $A_{1g}$ extended $s$-wave gap and for other regions a $B_{1g}$ $d$-wave gap is favored. In the following, we will examine the possibility of a resonance in the inelastic neutron scattering for each of these cases as well as for an $A_{1g}$ sign-switched $s$-wave gap.

The model that we will study consists of a 5-orbital ($d_{xz}$, $d_{yz}$, $d_{xy}$, $d_{x^2-y^2}$, $d_{3z^2-r^2}$) tight-binding fit to the DFT bandstructure of Cao {\it et al.}~\cite{ref:15} and onsite intra-orbital $U$, inter-orbital $V$ Coulomb and exchange $J$ interactions. The Hamiltonian is discussed in the Appendix and energies will be measured in units of the largest hopping matrix element $t^{11}_y$, i.e. the nearest-neighbor hopping along the $y$-direction between $d_{xz}$ orbitals. For this multiorbital problem, the dynamic spin susceptibility $\chi_{ij}(q,\omega)$ that determines the neutron scattering intensity depends upon an orbital-dependent spin susceptibility tensor~\cite{ref:16,ref:kar} given by 
\begin{equation}
	\chi^{rs}_{tu}(q,i\omega_m)=\int^\beta_0d\tau e^{i\omega_m\tau} \langle T_\tau S^{rs}_-(q,\tau)S^{tu}_+(-q,0)\rangle \label{eq:1} 
\end{equation}
Here $r$ and $s$ label orbital indices ($1,\dots,5$) corresponding to the $(d_{xz}$, $d_{yz}$, $d_{xy}$, $d_{x^2-y^2}$, $d_{3z^2-r^2}$) orbitals. $S^{rs}_+(q)=\frac{1}{2}\sum_kd^+_{r\uparrow}(k+q) \sigma^i_{\alpha\beta}d_{s\downarrow}(k)$ is the spin flip up operator acting between the $r$ and $s$ orbitals and $S^{sr}_-(-q)=[S^{rs}_+(q)]^\dagger$. Carrying out the usual analytic continuation of Matsubara frequencies to the real frequency axis, the dynamic spin susceptibility is given by 
\begin{equation}
	\chi(q,\omega)=\sum_{r,t}\chi^{rr}_{tt}(q,i\omega_m\to\omega+i\delta) \label{eq:2} 
\end{equation}

We will approximate the orbital spin susceptibility by an RPA-BCS form 
\begin{equation}
	\chi^{rr}_{tt}(q,\omega)=\sum_{r,t}\left\{ \chi_0(q,\omega)\left[1-U^s\chi_0(q,\omega)\right]^{-1}\right\}^{rr}_{tt} \label{eq:3} 
\end{equation}
with $U^s$ an interaction tensor given in the Appendix. This interaction tensor contains onsite intra- and inter-orbital Coulomb interaction $U$ and $V$ along with the intra-orbital exchange $J$ and pair hopping term $J'$.

Here $\chi_0$ is the BCS susceptibility tensor 
\begin{eqnarray}
	(\chi_0)^{rs}_{tu}(q,\omega_m)&=&-\frac{1}{2}\sum_{kn,\nu\nu'}M^{\nu\nu'}_{rstu}(k,q)\nonumber\\
	&\times&\bigl\{G^\nu(k+q,\omega_n+\omega_m)G^{\nu'}(k,\omega_n)\nonumber\\
	&+&F^\nu(-k-q,-\omega_n-\omega_m)F^{\nu'}(k,\omega_n)\bigr\}. \label{eq:5} 
\end{eqnarray}
with 
\begin{equation}
	G^\nu(k,\omega_n)=\frac{i\omega_n+E_\nu(k)}{\omega^2_n+{\cal E}^2_\nu(k)}, \hspace{5mm} F^\nu(k,\omega_n)=\frac{\Delta(k)}{\omega^2_n+{\cal E}^2_\nu(k)} \label{eq:6} 
\end{equation}
and ${\cal E}_\nu(k)=\sqrt{E^2_\nu(k)+\Delta^2(k)}$. The band energies $E_\nu(k)$ are measured relative to the Fermi energy and the orbital-band matrix elements $a^r_\nu(k)$ enter in determining 
\begin{equation}
	M^{\nu\nu'}_{rstu}(k,q)=a^{r^*}_\nu(k+q)a^s_{\nu'}(k)a^{t^*}_{\nu'}(k)a^u_\nu(k+q)\,. \label{eq:7} 
\end{equation}

\begin{figure}
	[htbp] 
	\includegraphics[width=9cm,clip,angle=0]{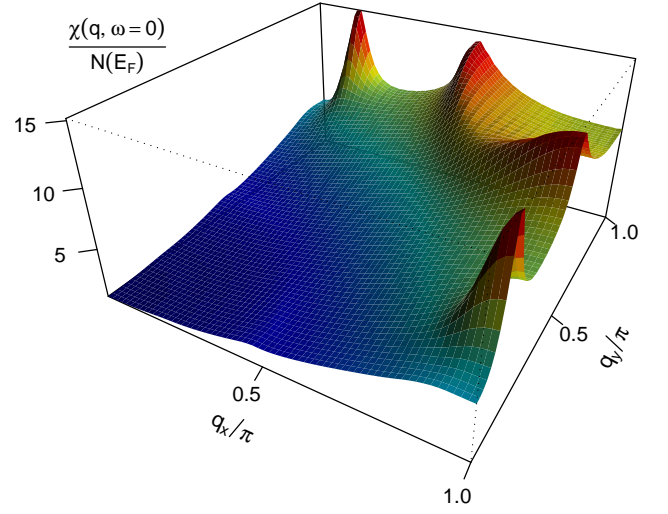} \vskip -1cm 
	\caption{The normal state static RPA spin susceptibility $\chi(q,0)$ normalized to the single particle density of states $N(E_F)$ per spin for a doping $x=0.125$ with $V=U-\frac{5}{4}J$, $J=U/8$ and $U=4$. The peaks are associated with scattering between the $\alpha_1$ and $\beta_1$ and $\beta_2$ Fermi surfaces shown in Fig.~\protect\ref{fig:1}.} 
	\label{fig:2} 
\end{figure}

Motivated by our previous calculations of the pairing interaction and gap functions for the 5-orbital model, we will examine the neutron scattering response for a typical set of interaction strengths $V=U-\frac{5}{4}J$, $J=U/8$, $J'=J/2$ with $U=4$ in units of $t^{11}_y$ at a doping $x=0.125$. The static RPA susceptibility $\chi(q,\omega=0)$ in the normal state is plotted in Fig.~\ref{fig:2}. 
Here the $(\pi,0)$ peak of the undoped system has moved to an incommensurate wave vector $q^*=(\pi,0.15\pi)$. This peak is associated with scattering between the $\alpha_1$ and $\beta_1$ Fermi surfaces. The ridge near $(\pi,\pi)$ is associated with scattering between the $\beta$ Fermi surfaces. The superconducting gap arising from the fluctuation-exchange pairing interaction for this case can be parameterized near the Fermi surfaces in terms of low order harmonics. For these interaction parameters, the leading pairing instability occurs in an $A_{1g}$ channel and the next leading instability is in a $B_{1g}$ channel. As discussed in Ref.~\onlinecite{ref:7}, the pairing strengths for these two gap symmetries are quite similar.

\begin{figure}
	[htbp] 
	\vskip -1.0cm 
	\includegraphics[width=3.5in]{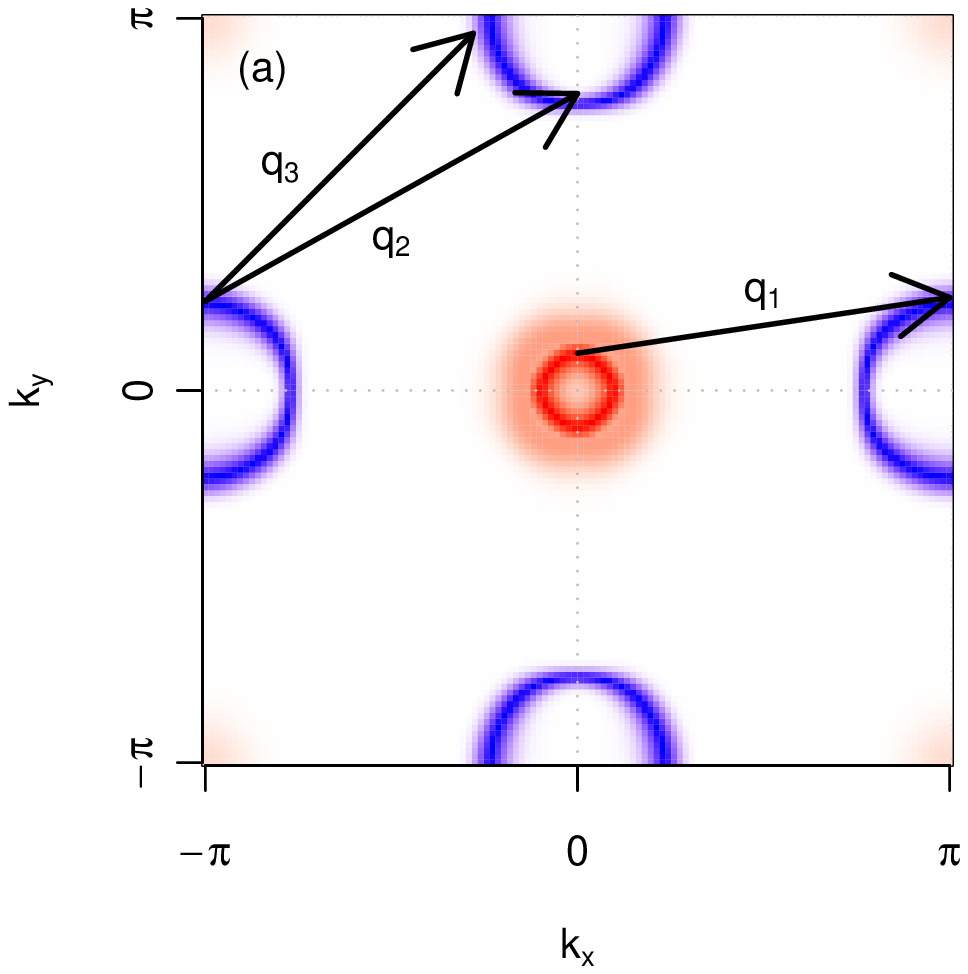} \vskip -2.2cm 
	\includegraphics[width=3.5in]{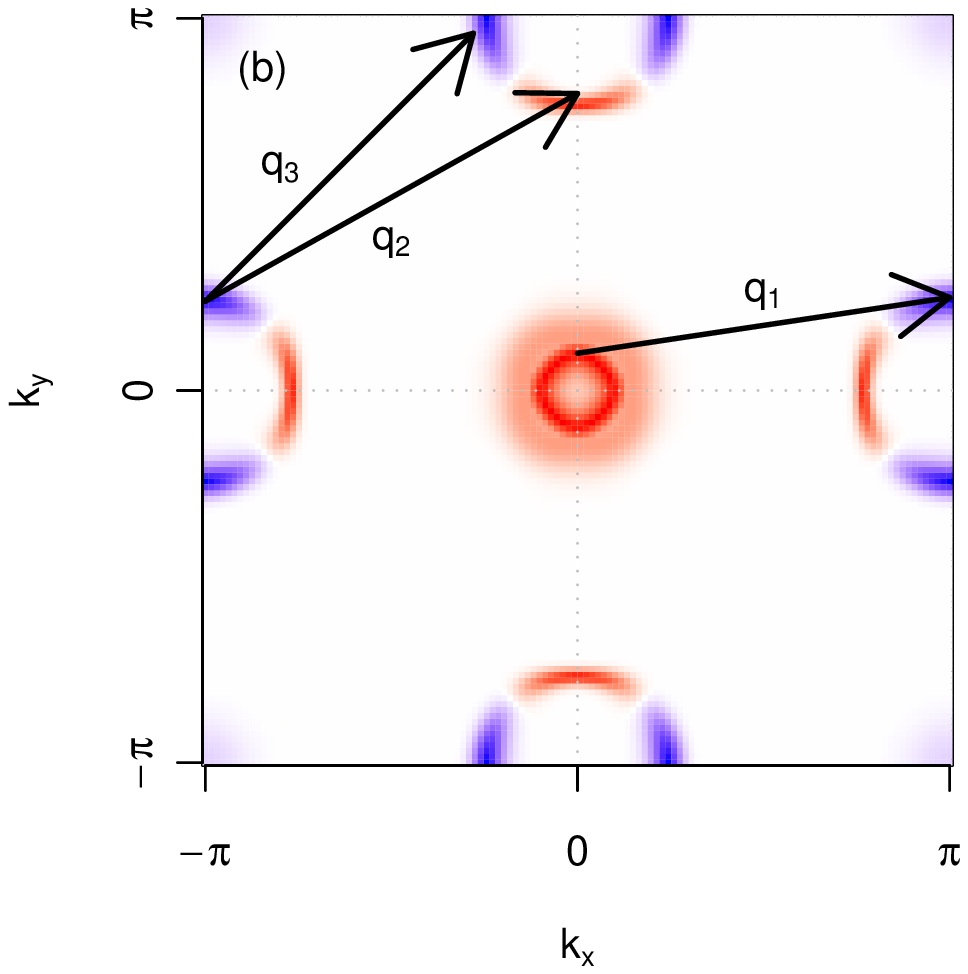} \vskip -2.2cm 
	\includegraphics[width=3.5in]{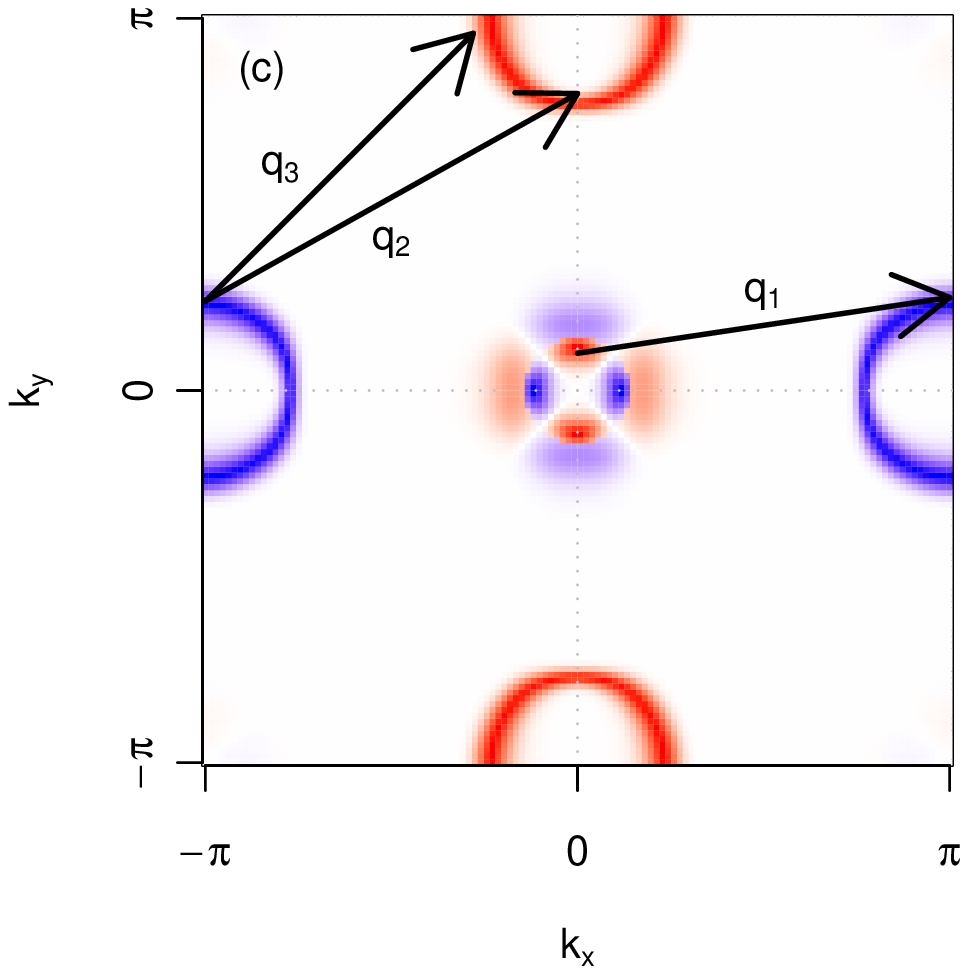}
	\caption{Plots of the gap $\Delta_\nu(k)$ for the $\alpha_1$, $\alpha_2$, and $\beta$ Fermi surfaces (a) the sign-switched $s$-wave, (b) the $A_{1g}$ (extended $s$-wave) gap $\Delta_\nu(\cos k_x+\cos k_y)$ and (c) the $B_{1g}$ ($d$-wave) gap $\Delta_\nu(\cos k_x-\cos k_y)$. All are plotted with the Gaussian cut-off. Here red denotes positive and blue negative values of the gap.} 
	\label{fig:3} 
\end{figure}

We will consider both of these possibilities using parameterized gaps given by 
\begin{equation}
	\Delta_\nu(k)=\Delta_\nu(\cos k_x\pm\cos k_y) \label{eq:8} 
\end{equation}
Here the plus sign corresponds to the $A_{1g}$ (extended $s$-wave) gap and the negative sign to the $B_{1g}$ ($d$-wave) gap. The amplitudes $\Delta_\nu$ are adjusted so that the maximum magnitude of the gap on the $\alpha_1$ and $\beta$ Fermi surfaces is 0.1 and 0.05 on the $\alpha_2$ Fermi surface. Based on the results from the fluctuation-exchange calculation, we have taken $\Delta_{\alpha_2}$ with the opposite sign to $\Delta_{\alpha_1}$ for the $B_{1g}$ case. We will also calculate the inelastic neutron scattering for the case of a sign-reversed $s$-wave system in which the gaps are isotropic on each Fermi surface with $\Delta_{\alpha_1}=0.1$, $\Delta_{\alpha_2}=0.05$ and $\Delta_{\beta_1} =\Delta_{\beta_2}=-0.1$. In each of these cases, in order to describe the behavior of the order parameter away from the Fermi surface, we have multiplied $\Delta_\nu(k)$ by a Gaussian cutoff $\exp[-(E(k)/\Delta E)^2]$ with $\Delta E$ of order several times the gap \cite{ref:7}. The resulting gaps are illustrated in Fig.~\ref{fig:3}. 
The anisotropic $A_{1g}$ s-wave gap structure in Fig.~\ref{fig:3}b is similar to that obtained in both RPA~\cite{ref:4,ref:7} and functional renormalization group calculations~\cite{ref:9}. Depending on the interaction and doping parameters, the variation of the gap on the $\beta$ Fermi surfaces may or may not be sufficient to produce nodes. The $d$-wave $B_{1g}$ gap illustrated in Fig.~\ref{fig:3}c is also similar to the gap structure found in these calculations. As noted, for the parameter set we have chosen it is the second leading instability but with other parameter choices it can become the leading instability.

\begin{figure}
	[htbp] \vskip -1.0cm 
	\includegraphics[width=3.5in,clip,angle=0]{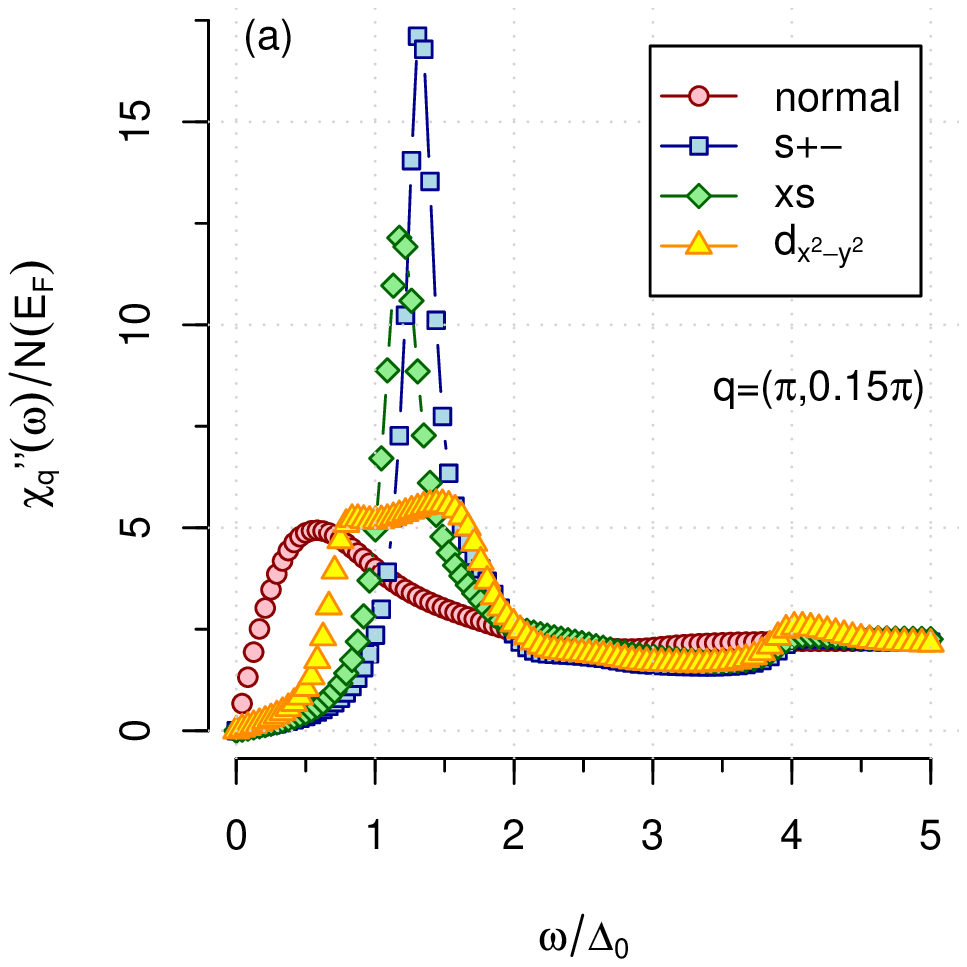} \vskip -1.9cm 
	\includegraphics[width=3.5in,clip,angle=0]{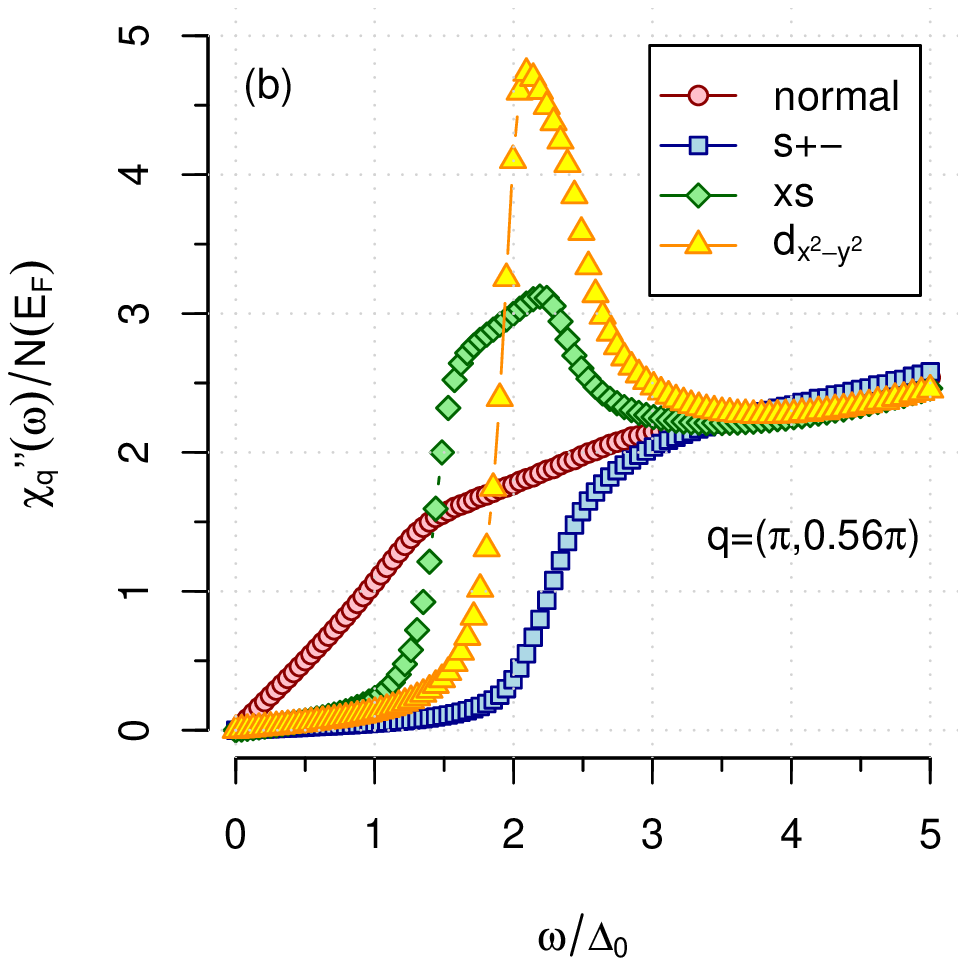} \vskip -1.9cm 
	\includegraphics[width=3.5in,clip,angle=0]{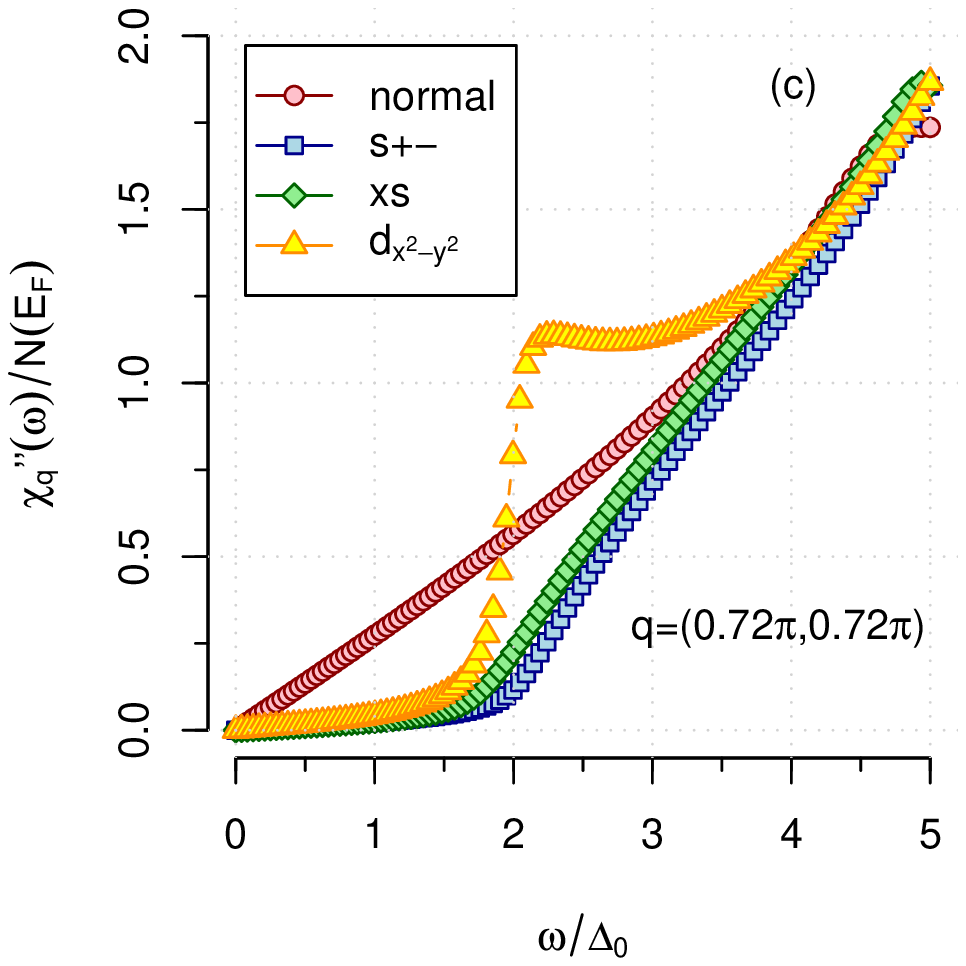} 
	\caption{The RPA-BCS dynamic spin susceptibility $\chi^{''}(q^*,\omega)$ versus $\omega$ for (a) $q^*=(\pi,0.15\pi)$, (b) $q^*=(\pi,0.56\pi)$ and (c) $q=(0.72\pi,0.72\pi)$ normalized to $N(E_F)$. The two $A_{1g}$ gaps exhibit a resonance and the $d_{x^2-y^2}$ gap has a broad response at $(\pi,0.15\pi)$. Only the $B_{1g}$ $d$-wave and the $A_{1g}$ extended $s$-wave show a response at $(\pi,0.56\pi)$ and at $(0.72\pi,0.72\pi)$ only the $d_{x^2-y^2}$ gap has a resonance.} 
	\label{fig:4} 
\end{figure}

The inelastic neutron scattering intensity is proportional to the imaginary part of $\chi(q,\omega)$. In Figure~\ref{fig:4} we show $\chi''(q,\omega)$ versus $\omega$ for the normal state and the three different superconducting gaps. 
These calculations have all been carried out at a temperature $T=0.005$ and the normal state has simply been included for comparison purposes. Figure~\ref{fig:4}a shows results for a momentum transfer $q_1=(\pi,0.15\pi)$, where we expect from Fig.~\ref{fig:1} that a resonance will appear. When the gap opens in the superconducting state, low energy spectral weight is shifted to higher energies and the quasi-particle-hole damping is suppressed. The resonance appears when the real part of the denominator of Eq.~\ref{eq:3} vanishes provided that $Re\Delta(k)\Delta(k+q^*)<0$ so that the coherence factor 
\begin{equation}
	\frac{1}{2}\left(1-\frac{\Delta(k)\Delta(k+q^*)}{{\cal E}(k){\cal E}(k+q^*)}\right) \label{eq:10} 
\end{equation}
goes to 1 rather than zero. Thus, just as in the well known cuprate case, the observation of a resonance provides evidence of a change in the relative signs of the gap between two regions of the Fermi surface \cite{ref:bulut,ref:eschrig}. Here for $q_1=(\pi,0.15\pi)$ shown in Fig.~\ref{fig:3}a, the resonance implies that this sign change occurs between regions on the $\alpha_1$ and $\beta$ Fermi surfaces. For the sign switched $s_\pm$ gap shown in Fig.~\ref{fig:3}a this sign change occurs over the entire Fermi suface. For the extended s-wave gap illustrated in Fig.~\ref{fig:3}b, the $q_1$ wavevector again dominantly connects regions of the $\alpha$ and $\beta$ Fermi sheets which have opposite signs. Because of the gap nodes and gap anisotropy on the $\beta$ Fermi surface sheet, the response is weaker than that for the sign-switched s-wave.

For $q_2=(\pi,0.56\pi)$ shown in Fig.~\ref{fig:3}b, there are important contributions involving particle-hole scattering between the $\beta_1$ and $\beta_2$ Fermi surfaces. In this case, the sign switched $s$-wave gap has the same sign on both of the $\beta$ Fermi surfaces so that the coherence factor, Eq.~\ref{eq:10} vanishes. Then as shown in Fig.~\ref{fig:4}b, there is no resonance for the $s^\pm$ gap and one only sees that spectral weight is shifted from low frequencies in the normal state to frequencies above $\sim2\Delta$ in the superconducting state. However, the resonance is present for both the extended $s$-wave state and the $B_{1g}$ ($d$-wave) state since, as seen in Fig.~\ref{fig:3}, $q_2$ connects regions of the $\beta_1$ and $\beta_2$ Fermi surfaces where there is a sign change of the gap and $Re\Delta(k)\Delta(k+q^*)<0$. Again these regions are smaller for the extended s-wave, Fig.~\ref{fig:3}b, compared to the $d_{x^2-y^2}$ gap shown in Fig.~\ref{fig:3}c.

For the $q_3=(0.72\pi,0.72\pi)$ momentum transfer shown in Fig.~\ref{fig:3}b, only the $d_{x^2-y^2}$ gap leads to a resonance response. In this case, as in the previous case, the scattering involves transitions between the $\beta$ sheets. For the $d_{x^2-y^2}$-wave gap, Fig.~\ref{fig:3}c, these sheets have opposite signs, but for the extended s-wave, Fig.~\ref{fig:3}b, these regions have the same sign and the resonance is suppressed.  

These calculations show that both the sign-switched s-wave and the extended s-wave gaps are consistent with the occurrence of a resonance in the neutron scattering response which is observed in the superconducting state of the 122 Fe pnictides near $(\pi,0)$ in the unfolded (1-Fe/zone) Brillouin zone. In our calculations, this resonance is strongest for the phenomological, sign-switched gap. However it also appears as a clear resonance for an extended s-wave gap and even as a weak broad structure for a $d_{x^2-y^2}$ gap. At a momentum transfer $q_2=(\pi,0.56\pi)$, the transitions involve scattering between regions of the $\beta_1$ and $\beta_2$ Fermi surfaces where the sign-switched s-wave gap has the same sign.  For the diagonal momentum transfer shown in Fig.~\ref{fig:4}c, only the d-wave gap exhibits a resonance. Thus further neutron measurements at other momentum transfers can narrow the possible gap structure. Our results suggest that cuts in q-space going from $(\pi,0)$ to $(\pi,\pi)$  and along the diagonal direction in the unfolded 1-Fe zone are most promising in providing useful information.

\section*{Acknowledgements} TAM and DJS would like to acknowledge useful discussions with A.D.~Christianson, M.~Lumsden, D.~Mandrus and H.A.~Mook. They would also like to acknowledge support from Oak Ridge National Laboratory's Center for Nanophase Materials Sciences and the Scientific User Facilities Division, Office of Basic Energy Sciences, U.S.~Department of Energy. PJH would like to acknowledge support from the DOE grant DOE DE-FG02-05ER46236.

\section{Appendix} \setcounter{equation}{0} 
\renewcommand{\theequation}{A-\arabic{equation}} The full DFT band structure can be fitted in the vicinity of the Fermi energy using a tight-binding approximation with the five Fe $d$ orbitals as basis set. Here we introduce a coordinate system aligned parallel to the nearest neighbor Fe-Fe direction. The Hamiltonian for this 5 orbital model can be written as 
\begin{equation}
	H_0 = \sum_{k,\sigma} \sum_{mn} (\xi_{mn}(k) + \epsilon_m \delta_{mn}) d_{m\sigma}^\dagger (k) d_{n\sigma} (k) \label{eq:A1} 
\end{equation}
where $d_{m\sigma}^\dagger (k)$ creates a particle with momentum $k$ and spin $\sigma$ in orbital $m$. The symmetry of the kinetic energy terms $\xi_{mn}(k)$ can be derived from a Slater-Koster based parametrization that respects the symmetry of the FeAs layers. The exact sizes of the hopping parameters can be used as fitting parameters to approximately reproduce the band energies at the high symmetry points ($\Gamma$, X, M) and also the band structure along the high symmetry directions -- at least in the vicinity of the Fermi energy. The explicit form of the kinetic energy terms $\xi_{mn}$ as well as the on-site energies $\epsilon_m$ and the values of the hopping terms used for the fitting of the band structure by Cao {\it et al.}~\cite{ref:15} can be found in the appendix of Ref.~9.

%\onlinecite{ref:7}%.
The interaction tensor $U^s$ is given by 
\begin{equation}
	(U^s)^{aa}_{aa}=U\,, (U^s)^{aa}_{bb}=J/2\,, (U^s)^{ab}_{ab}=\frac{J}{4}+V\,, (U^s)^{ba}_{ab}=J' \label{eq:A2} 
\end{equation}
where $a\ne b$.

\end{document}